\documentclass[twocolumn]{revtex4-1}
\usepackage{amsmath}
\usepackage{tipa}
\usepackage{amssymb}
\usepackage{graphicx}
\usepackage{xfrac}
\usepackage{bbm}
\usepackage{enumerate}
\usepackage{color}
\usepackage[colorlinks = true,
            linkcolor = blue,
            urlcolor  = blue,
            citecolor = blue,
            anchorcolor = blue]{hyperref}
\usepackage{subcaption}
\usepackage{physics}
\usepackage{rotating}
\usepackage{capt-of}
\usepackage{chngcntr}
\usepackage{varwidth}
\usepackage{slashed}

\newcommand{\eq}[1]{\begin{equation}#1\end{equation}}
\newcommand{\al}[1]{\begin{align}#1\end{align}}
\newcommand{\nn}{\nonumber}

\def\hri#1#2{\href{http://arxiv.org/abs/#1}{[arXiv:#1[#2]]}}
\def\hre#1#2{\href{http://arxiv.org/abs/#1/#2}{[arXiv:#1/#2]}}
\def\doi#1#2{\href{https://doi.org/#2}{#1}}

\newcommand{\be}{\begin{equation}}
\newcommand{\ee}{\end{equation}}

\begin{document}

\title{Chiral symmetry breaking and monopoles in gauge theories}

\author{Adith Ramamurti}
\email[]{adith.ramamurti@stonybrook.edu}
\author{Edward Shuryak}
\email[]{edward.shuryak@stonybrook.edu}

\affiliation{Department of Physics and Astronomy, \\ Stony Brook University,\\
Stony Brook, NY 11794, USA}

\date{\today}

\begin{abstract}

QCD monopoles are magnetically charged quasiparticles whose Bose-Einstein condensation (BEC) at $T<T_c$ creates electric confinement and flux tubes.  The ``magnetic scenario" of QCD proposes that scattering on the non-condensed component of the monopole ensemble at $T>T_c$ plays an important role in explaining the properties of strongly coupled quark-gluon plasma (sQGP) near the deconfinement temperature. In this paper, we study the phenomenon of chiral symmetry breaking and its relation to magnetic monopoles. Specifically, we study the eigenvalue spectrum of the Dirac operator in the basis of fermionic zero modes in an SU(2) monopole background. We find that as the temperature approaches the deconfinement temperature $T_c$ from above, the eigenvalue spectrum has a finite density at $\omega = 0$, indicating the presence of a chiral condensate. In addition, we find the critical scaling of the eigenvalue gap to be consistent with that of the correlation length in the 3d Ising model and the BEC transition of monopoles on the lattice.

\end{abstract}
\maketitle

\section{Introduction}
The possible existence of magnetic monopoles in electrodynamics fascinated leading physicists in the 19th century. With the development of quantum mechanics, Dirac famously related the existence of monopoles with electric charge quantization \cite{Dirac60}. Monopoles in quantum electrodynamics, however, were never found.
  
With the advent of non-Abelian gauge theories, classical solitons with magnetic charge were found by 't Hooft \cite{NUPHA.B79.276} and Polyakov \cite{Polyakov:1974ek} in the Georgi-Glashow model. Such monopoles play important role in all other theories with an adjoint scalar field, notably in theories with extended supersymmetry ${\cal N} = 2,\,4$.

In theories without such scalars, e.g. pure gauge theories, one can use the same 't Hooft-Polyakov solution with the zeroth component of the gauge field $A_0$ acting as the ``Higgs;" this option leads to the semiclassical theory of instanton-dyons. For a recent short review, see Ref. \cite{Shuryak:2017kct}. These objects, however, are {\em pseudo}-particles and not particles, existing only in the Euclidean formulation of the theory, which severely limits their phenomenological applications. We will not discuss this issue further and only remind the reader that, in the case of extended supersymmetry, the partition functions in terms of monopoles and instanton-dyons were shown to be equal, related by the so-called Poisson duality \cite{Dorey:2000qc,Poppitz:2011wy,Poppitz:2012sw}.

In spite of the monopole solution lacking in theories without scalars,  Nambu \cite{Nambu:1974zg}, 't Hooft \cite{NUPHA.B190.455}, and Mandelstam \cite{Mandelstam:1974pi} proposed the ``dual superconductor" model of the electric color confinement. In this model, the Bose-Einstein condensation (BEC) of monopoles at $T\leq T_c$ expels electric field from the vacuum into confining flux tubes. 

In lattice studies of gauge theories, monopoles have been identified via procedures including choosing specific gauges, such as Maximal Abelian Gauge. Their locations and paths were positively correlated with gauge-invariant observables, such as the action and square of the magnetic field \cite{Laursen:1987eb}.  The monopoles were found to create a magnetic current around the electric flux tube, stabilizing them \cite{Koma:2003gq, Bornyakov:2003vx}. In the Landau gauge, while monopole-type singularities themselves are not present, the physical properties that they source, e.g. the magnetic displacement current, are still present and are gauge-invariant \cite{Suzuki:2009xy}. Furthermore, their motion and correlations were shown to be exactly as expected for a Coulomb plasma \cite{DAlessandro:2007lae, Bonati:2013bga, Liao:2008jg}. The deconfinement critical temperature $T_c$  does coincide accurately with that of monopole BEC transition \cite{D'Alessandro:2010xg,Bonati:2013bga,Ramamurti:2017fdn}, and the BEC transition, along with the the magnetic charge, has been shown to be gauge independent \cite{Bonati:2010tz, Bonati:2010bb, DiGiacomo:2017blx}.
 
The ``magnetic scenario" of QCD \cite{Liao:2006ry,Liao:2007mj,Liao:2008jg} assumes the presence of non-condensed monopoles in quark-gluon plasma (QGP). Unlike quarks and gluons, which have vanishing densities at  $T\rightarrow T_c$, the monopole density has a peak near $T_c$ and is dominant there. Monopole-gluon and monopole-quark scattering were shown to play a significant role in kinetic properties of QGP, such as the shear viscosity $\eta$ \cite{Ratti:2008jz} and the jet quenching parameter $\hat q$  \cite{Xu:2015bbz,Xu:2014tda,Ramamurti:2017zjn}. The non-condensed monopoles should also lead to electric flux tubes at  $T>T_c$  \cite{Liao:2007mj}, which indeed were recently observed on the lattice \cite{Cea:2017ocq}. Thus, there is a growing amount of phenomenological evidence suggesting magnetic monopoles do exist, not only as a confining condensate at $T\leq T_c$, but also as non-condensed quasiparticles at $T>T_c$.

While the central role of monopoles in the confinement-deconfinement transition was recognized long ago, their relation to another important non-perturbative aspect of QCD-like theories, chiral symmetry breaking, has attracted much less attention. It has been found on the lattice that, by decomposing the gauge fields into Abelian-monopole, Abelian-plain, and non-Abelian components, the removal of the monopoles does indeed lead to removal of the quark condensate \cite{Suzuki:1989gp, Miyamura:1995xn}. 

In this paper, we address how chiral symmetry breaking and the generation of the non-zero quark condensate at $T<T_c$ can be explained in terms of this monopole model. The mechanism is based on the formation of bound states between quarks and monopoles. Like in condensed matter systems, in which ``doping" of a material by atoms with an extra state leads to new set of states and alters its conductivity, the presence of monopoles radically affects the Dirac eigenvalue spectrum.  

One obvious difficulty of the problem is the fact that a detailed understanding of the  ``lattice monopoles" is lacking; they are treated as effective objects whose parameters and behavior we can observe on the lattice and parameterize, but their microscopic structure has yet to be understood. In particular, the 't Hooft-Polyakov monopole solution includes a  chiral-symmetry-breaking scalar field, while we know that, in massless QCD-like theories, chiral symmetry is locally unbroken. We assume that the zero modes in question are chiral themselves, like they are in the instanton-dyon theory, and that chiral symmetry breaking can only be achieved by a spontaneous breaking of the symmetry.      

The other difficulty of the problem is the important distinction between fermionic zero modes of (i) monopoles and (ii) instanton-dyons. The latter include the so called $L$-type or twisted dyons, which possess 4-dimensional zero modes for antisymmetric fermions. Therefore, their collectivization naturally leads to chiral symmetry breaking, studied recently in Ref. \cite{Larsen:2015tso}, in a natural mechanism originally proposed for the instantons; for a review, see Ref. \cite{Schafer:1996wv}. As follows from Banks-Casher relation \cite{Banks:1979yr}, the quark condensate is proportional to density of Dirac eigenstates at zero eigenvalue.
 
The monopoles also have fermionic zero modes \cite{Jackiw:1975fn}, which are 3-dimensional. They are, therefore, simply a bound state of a fermion and a monopole. In theories with extended supersymmetries, such objects do exist, fulfilling an important general requirement that monopoles need to come in particular super-multiplets, with fermionic spin 1/2 for ${\cal N}=2$, or spins 1/2 and 1 for ${\cal N}=4$. The anti-periodic boundary conditions for fermions in Matsubara time implies certain time dependence of the quark fields, and (as we will discuss in detail below) the lowest 4-dimensional Dirac eigenvalues produced by quarks bound to monopoles are the values $\lambda =\pm \pi T$, not at zero. 

This, however, is only true for a single monopole. In a monopole {\em ensemble} with non-zero density, the monopole-quark bound states are collectivized and Dirac eigenvalue spectrum is modified. The question is whether this effect can lead to a nonzero $\rho(\lambda=0)\propto \langle \bar q q \rangle $, and if so, whether it happens at the temperature at which chiral symmetry breaking is observed. As we will show below, we find affirmative answers to both these questions. The phenomenological monopole model parameters are such that a non-zero quark condensate is generated by monopoles at $T\approx T_c$. 

As a final introductory comment, we note that our approach is to take as inputs the empirical monopole density $n(T)$ and ensemble of paths from our previous study. Using them, we calculate the corresponding Dirac eigenvalue spectrum. The back-reaction of the fermions on monopole density and their motion is neglected. In this respect, our calculation is ideological similar to {\em quenched} lattice calculations, which also ignore quark back-reaction on the gauge fields.

\section{Fermionic zero modes of SU(2) monopoles}

In this section, we will overview the calculation of the fermion-monopole zero modes in the Georgi-Glashow model. First we will remind the reader of the result found by Jackiw and Rebbi \cite{Jackiw:1975fn} for the fermionic zero modes, and then go on to compute the matrix element between two monopoles in the basis of zero modes.

\subsection{Fermionic zero mode of a monopole }

This section is introductory,  summarizing  known results from Refs. \cite{Jackiw:1975fn,Callias:1977cc} and presented for self-completeness of the paper.  The Lagrangian of  the  Georgi-Glashow model (without fermions)  is  
\eq{
\mathcal{L} = -\frac{1}{4} F^{\mu\nu}_a F_{a\mu\nu} + \frac{1}{2}(D_\mu \phi)_a (D^\mu \phi)_a - \frac{1}{4} \lambda(\phi_a\phi_a-v^2)\,
}
where 
\eq{F^{\mu\nu}_a = \partial^\mu A^\nu_a - \partial^\nu A^\mu_a + g \epsilon_{abc}A^\mu_b A^\nu_c\,, \nn}
and
\eq{(D^\mu \phi)_a = \partial^\mu\phi_a + g \epsilon_{abc}A^\mu_b \phi_c\,. \nn}

The fermionic part considered by Jackiw and Rebbi is
\eq{
\mathcal{L_F} = i \bar{\psi}_n \gamma^\mu (D_\mu \psi)_n - G g \bar{\psi}_n \tau^a_{nm}\psi_m\phi_a\,, 
}
with $G$ a constant, $\tau^a = \sigma^a/2$, and 
\eq{(D_\mu \psi)_n = \partial^\mu\psi_n-ig\tau^a_{nm}A^\mu_a\psi_m\,. \nn}

The 't Hooft-Polyakov  
monopole solution has the form
\al{
A^0_a &= 0\,, \nn \\
A^i_a &= \epsilon^{aij} \hat{r}_j \frac{A(r)}{g}\,, \\
\phi_a &= \hat{r}_a \frac{\phi(r)}{g}\,. \nn
}
With this ansatz, the equations of motion from the pure-gauge Lagrangian are
\al{
0 = \frac{2}{r^2}\frac{\dd}{\dd r}& \left( r^2 \frac{\dd A}{\dd r}\right) - \frac{2}{r}\frac{\dd A}{\dd r} + \frac{2}{r^2}\frac{\dd}{\dd r}(rA) \nn \\
& -\frac{6}{r^2}A-\frac{6g}{r}A^2 - 2g^2A^3 - \phi\left(\frac{2g}{r} + 2g^2A\right)\,, 
}
\al{
0 = \frac{1}{r^2}\frac{\dd}{\dd r} \left( r^2 \frac{\dd \phi}{\dd r}\right)& - \frac{2}{r^2} \phi - \frac{4g}{r} A \phi \nn \\
&-2g^2A^2\phi - 2U'(|\phi|^2)\phi \,, \nn
}
and boundary conditions,
\eq{
\left(r^2 \frac{\dd A}{\dd r} + 2rA\right)\Bigg|_{r=0} = 0\,,
}
\eq{
\left(r^2 \frac{\dd \phi}{\dd r}\right)\Bigg|_{r=0} = 0\,. \nn
}

This set of equations and boundary conditions, combined with single-valuedness of the fields at the origin give
\eq{
A(0) = \phi(0) = 0\,, \nn
}
\eq{
A(r)\Big|_{r\rightarrow\infty} = -\frac{1}{r}\,, 
}
\eq{
\phi(r)\Big|_{r\rightarrow\infty} = v\,, \nn
}
with $v$ a constant.

The Dirac equation for the fermion field is written in the form 
\al{
\Big[-i\vec\alpha\cdot\vec\partial\delta_{nm}&+\frac{1}{2}A(r)\sigma^a_{nm}(\vec\alpha\times\vec{\hat{r}})_a  \nn \\ 
&+ \frac{G \phi(r)}{2} \sigma^a_{nm}\hat r_a\beta\Big] \psi_m = E \psi_n\,,
}
where $n,m=1,2$ are the isospin indices, $\sigma^a$ are the Pauli matrices, and
  \begin{align}
    \alpha_i  =   \begin{pmatrix}
          0 & \sigma_i \\
           \sigma_i & 0\\
         \end{pmatrix} \,, \hspace{1cm} 
          \beta  =  -i \begin{pmatrix}
          0 & \mathbbm{1}\\
           -\mathbbm{1} & 0\\
         \end{pmatrix} \,.
  \end{align}
For clarity, in Appendix \ref{app_a} we will discuss their representation of the Dirac matrices and the relation to fermion chirality.    
  
In this representation of the Dirac matrices, the $\alpha_i,\beta$ are non-diagonal,  and the only diagonal term is the energy. For a zero mode $E=0$, the problem is ``chiral," in the sense that the 4-spinor $\psi$ splits into separate upper and lower components,
  \begin{align}
    \psi  &= \begin{pmatrix}
          \psi^+ \\
           \psi^-\\
         \end{pmatrix} \,, \nn
  \end{align}
 where $\psi^\pm_{lm}$ has four components, with $l$ corresponding to spin indices and $m$ corresponding to isospin (The SU(2) color is called isospin in Georgi-Glashow model.) These upper and lower components can be further written as two scalar and vector fields,
 \eq{
 \psi^\pm_{lm} = (g^\pm\delta_{lm}+\vec{g}^\pm\cdot\vec\sigma_{ln})\sigma_{nm}^2\,. \nn
}

Carrying out the partial wave analysis (see Appendix of Ref. \cite{Jackiw:1975fn}) and finding the zero energy solution gives
\al{
\vec g^\pm(r) &= 0, \nn \\
g^-(r) &= c^- \times \exp \left(\int_0^r \dd r^\prime\left[A(r^\prime) + \frac{1}{2}G\phi(r^\prime)\right]\right)\,,
\nn \\
g^+(r) &= c^+ \times \exp \left(\int_0^r \dd r^\prime\left[A(r^\prime) - \frac{1}{2}G\phi(r^\prime)\right]\right)\,, \nn
}
The $g^-$ solution is un-normalizable, so it is set to zero ($c^-$ = 0). This gives the spinors
\al{
\psi^-_{lm} = &\,0\,, \nn \\
\psi^+_{lm} =&\,N\exp \left(\int_0^r \dd r^\prime\left[A(r^\prime) - \frac{1}{2}G\phi(r^\prime)\right]\right) \\&\nn \times (s^+_l s^-_m- s^-_l s^+_m)\,,
}
where $s^\pm$ are the eigenvectors of $\sigma^3$, and $N$ is a normalization.

The extension of the SU(2) 't Hooft-Polyakov monopole solution to SU(3) -- with the same Georgi-Glashow-like Lagrangian --  is discussed by A. Sinha \cite{Sinha:1976bw}. In QCD, the $A_0$ field plays the role of the scalar (Higgs) field in the Georgi-Glashow model. For the purposes of this work, and for simplicity, we will study only the SU(2) case.

\subsection{Quark hopping matrix}

Recognizing fermionic binding to monopoles, we now proceed to description of their dynamics in the presence of ensembles of monopoles. The basis of the description is assumed to be the set of zero modes described in the previous section. The Dirac operator is written as a  matrix in this basis, so that $i-j$ element is related to ``hopping" between them. Such an approach originated from the ``instanton liquid" model \cite{Schafer:1996wv}.  

The matrix elements of the ``hopping matrix''
\eq{
\bf{T} = \begin{pmatrix}
          0 & iT_{ij} \\
	iT_{ji} &0
         \end{pmatrix}
} where the $T_{ij}$s are defined as the matrix element,
\eq{
T_{ij} \equiv \bra{i} -i\slashed{D} \ket{j},
}
between the zero modes located on monopoles $i$ and antimonopoles $j$. In the SU(2) case we are considering, this is equivalent to
\al{
 T_{ij} = &\braket{\psi_i}{x}\bra{x}-i\slashed{D}\ket{y}\braket{y}{\psi_j} \nn \\
 = &\int \dd^3 x \psi_{kn}^\dagger(x-x_i) (-i \slashed{D}) \psi_{lm}(x-x_j) \nn\\
 =  &\int \dd^3 x \psi_{kn}^\dagger(x-x_i) \Big[-i(\vec\alpha\cdot\vec\partial +\vec \alpha\cdot\vec\partial  -\vec \alpha\cdot\vec\partial ) \nn\delta_{nm}\\ 
\nn  &+\frac{1}{2}(A(x-x_i)+A(x-x_j))\sigma^a_{nm}(\vec\alpha\times\vec{\hat{r}})_a 
\\ &+ \frac{G (\phi(x-x_i)+\phi(x-x_j))}{2} \sigma^a_{nm}\hat r_a\beta\Big]  \psi_{lm}(x-x_j) \nn \\
=&\int \sum_m \dd^3 x \psi_{km}^\dagger(x-x_i) [-i\vec \alpha\cdot\vec\partial ]^{kl}\psi_{lm}(x-x_j)
}
where $\psi$s are zero modes with origin at $x_{i,j}$, the locations of the two monopoles, $n,m$ are the isospin/color indices, and we have used the fact that applying the Dirac operator to these wavefunctions gives zero.

The operator between the wavefunctions is
\al{
i\vec \alpha\cdot\vec\partial= i 
	\begin{pmatrix}
          0 & 0 & 0 & 1 \\
           0 & 0 & 1 & 0\\
           0 & 1& 0 & 0 \\
           1 & 0 & 0 & 0\\
         \end{pmatrix}& \partial_x  +
         i \begin{pmatrix}
          0 & 0 & 0 & -i \\
           0 & 0 & i & 0\\
           0 & -i& 0 & 0 \\
           i & 0 & 0 & 0\\
         \end{pmatrix} \partial_y \nn \\& + 
        i  \begin{pmatrix}
          0 & 0 & 1 & 0 \\
           0 & 0 & 0 & -1\\
           1 & 0& 0 & 0 \\
           0 & -1 & 0 & 0\\
         \end{pmatrix} \partial_z \,.
}
In the case where the original wavefunction only has an upper component, the resulting vector after applying this operator has only the lower component. Therefore, the matrix elements of the ``hopping matrix'' $T_{ij}$ are zero unless the fermionic zero modes have {\em opposite} chirality. So, a left-handed fermion zero mode has non-zero overlap with a right-handed zero mode, and vice versa. To get the opposite chirality, we need to change the sign of the couplings. In the case of an SU(2) antimonopole, only the lower spinor survives, with the same wavefunction (the spin of the fermion flips). We will call these zero modes $\xi$.

This allows us to split the previous equation into two equations,
\al{
T_{ij} = \int \dd^3 x \xi^\dagger(x-x_i) [-i\vec \sigma \cdot\vec\partial ]\psi(x-x_j)\,, \nn \\
T_{ji} = \int \dd^3 x \psi^\dagger(x-x_i) [-i\vec \sigma \cdot\vec\partial ]\xi(x-x_j)\,,
}
where we use the appropriate one depending on whether $i(j)$ is a location of a monopole(antimonopole).

For the matrix element between a right-handed monopole zero mode and a left-handed anti-monopole zero mode,
\al{
T_{ij} = &\int \dd^3 \vec x \sum_m \xi_{km}^\dagger(\vec x-\vec x_i) [-i\vec \sigma \cdot\vec\partial ]^{kl}\psi_{lm}(\vec x-\vec x_j) \nn\,,}
where $m$ is the traced-over isospin/color index, and $k,l$ are the spin indices (which are all convoluted). The operator is,
\eq{
[-i\vec \sigma \cdot\vec\partial ] = \begin{pmatrix}
	-i \partial_z & -i \partial_x-\partial_y \\
	-i \partial_x+\partial_y & i\partial_z
         \end{pmatrix}\,.
}
For simplicity, we will treat each isospin/color case separately, so the only indices left are spin indices. We will denote the spatial-only part of the wavefunction (without spinors) with tildes.
\vspace{3ex}
\paragraph[]{\underline{m=1}:} \hspace{0pt} \\ 

We will write $\xi_{k1}$ as $\tilde\xi a_k$ and $\psi_{l1}$ as $\tilde\psi a_l$. The wavefunction has spinors
\eq{a_i \equiv s^+_i s^-_1- s^-_i s^+_1 = \begin{pmatrix} 0 \\ -1 \end{pmatrix}\,, \nn}
\eq{a_i^\dagger \equiv s^+_1 s^-_i- s^-_1 s^+_i = \begin{pmatrix} 1 & 0 \end{pmatrix}\,, \nn}
and so
\eq{
\tilde{\xi} a_k^\dagger  [-i\vec \sigma \cdot\vec\partial ]^{kl} a_l \tilde{\psi} = \tilde{\xi} (i\partial_x+\partial_y) \tilde{\psi}\,.
}

\paragraph[]{\underline{m=2}:} \hspace{0pt} \\ 

The wavefunction has spinors
\eq{b_i \equiv s^+_i s^-_2- s^-_i s^+_2 = \begin{pmatrix} 1 \\ 0  \end{pmatrix}\,, \nn}

\eq{b_i^\dagger \equiv s^+_2 s^-_i- s^-_2 s^+_i = \begin{pmatrix} 0 & -1  \end{pmatrix}\,,\nn}
and so
\eq{
\tilde{\xi} b_k  [-i\vec \sigma \cdot\vec\partial ]^{kl} b_l \tilde{\psi} = \tilde{\xi} (i\partial_x-\partial_y) \tilde{\psi}\,.
}

\vspace{3ex}
\noindent Combining the two cases,
\eq{
 \sum_m \xi_{km}^\dagger[-i\vec \sigma \cdot\vec\partial ]^{kl}\psi_{lm}  = 2i \tilde{\xi} \partial_x \tilde{\psi}\,,
 }
which then yields that,
\eq{
T_{ij} = 2i\int \dd^3 \vec x \tilde{\xi}(\vec x - \vec x_i) \partial_x \tilde{\psi}(\vec x- \vec x_j)\,.
}
For the matrix element between a left-handed anti-monopole zero mode and a right-handed monopole zero mode, we get, similarly,
\eq{
T_{ji} = 2i\int \dd^3 \vec x \tilde{\psi}(\vec x - \vec x_i) \partial_x \tilde{\xi}(\vec x- \vec x_j)\,. \nn
}

The full solutions to the equations of motion for $A(r)$ and $\phi(r)$ are well behaved at the origin, so we must use those in lieu of only considering asymptotics. The only analytic solution to the equations of motion is in the case where $\lambda=0$; this solution is known as the Bogomolnyi-Prasad-Sommerfeld (BPS) monopole \cite{Bogomolny:1975de,Prasad:1975kr}. In this case, 
\eq{
H(\zeta) = \zeta \coth(\zeta)-1\,, \nn
}
\eq{
K(\zeta) = \frac{\zeta}{\sinh(\zeta)}\,,
}
where $\zeta = gvr$. In terms of these functions, our gauge fields are,
\al{
A^0_a &= 0\,, \nn \\
A^i_a &= \epsilon^{aij} \frac{r_j}{g r^2}(1-K(\zeta))\,, \\
\phi_a &= \frac{r_a}{g r^2} H(\zeta)\,. \nn
}
This leads to the identification with our earlier notation,
\al{
A(r) &=  \frac{1-K(g v r)}{r}\,, \\
\phi(r) &= \frac{H(g v r)}{r}\,. \nn
}

For the monopole zero mode we get that, up to normalization,
\eq{
\tilde \psi =\frac{1}{2} (g v r)^{\frac{G}{2}+1} \coth \left(\frac{g v r}{2}\right) \sinh ^{-\frac{G}{2}}(g v r)\,.
}
Then, 
\al{
\partial_x \tilde \psi &=  -  \left(\frac{x}{4r^2} \right)(g v r)^{\frac{G}{2} + 1} \coth \left(\frac{1}{2} g v r\right) \sinh
   ^{-\frac{G}{2}-1}\left(g v r\right) \nn \\ &\times \left(-(G+2)  \sinh \left(g v r\right)+g G v
   r \cosh \left(g v r\right)+2 g v r\right)\,,
}
and similarly for the anti-monopole zero mode wave function. Putting these solutions into the hopping matrix element equation for monopole-to-antimonopole, we get
\al{
 T_{ij}(r_0) =& 2i  \int \dd^3 \vec x \tilde{\xi}(|r-r_0|) \partial_x \tilde{\psi}(r) \nn \\
 \nn =  & 2i N^2 \int \dd^3 \vec x \left(-\frac{x}{8r^2}\right) (g v)^{G+2} r^{\frac{G+2}{2}}|r-r_0|^{\frac{G+2}{2}}
 \\ \nn & \times \coth \left(\frac{1}{2} g v r\right)  \sinh
   ^{-\frac{G}{2}-1}\left(g v r\right) \nn 
   \\ \nn  &  \times \coth \left(\frac{1}{2} g v
   |r-r_0|\right) \sinh ^{-\frac{G}{2}}\left(g v
   |r-r_0|\right)\\ & \times 
   \left(-(G+2)  \sinh \left(g v r\right) +g G v r \cosh \left(g v
   r\right)+2 g v r\right) \,.
   \label{eq_tij}
   }

The  combination $gvr$ is dimensionless, as is $x/r$, so the integrand has dimension [energy] ($1/r$). The parameter $v$, in the BPS limit, is determined by the mass of the monopole, and $g$ is taken to be the same as in the strong coupling constant.

\begin{figure}
\begin{center}
\includegraphics[width=.5\textwidth]{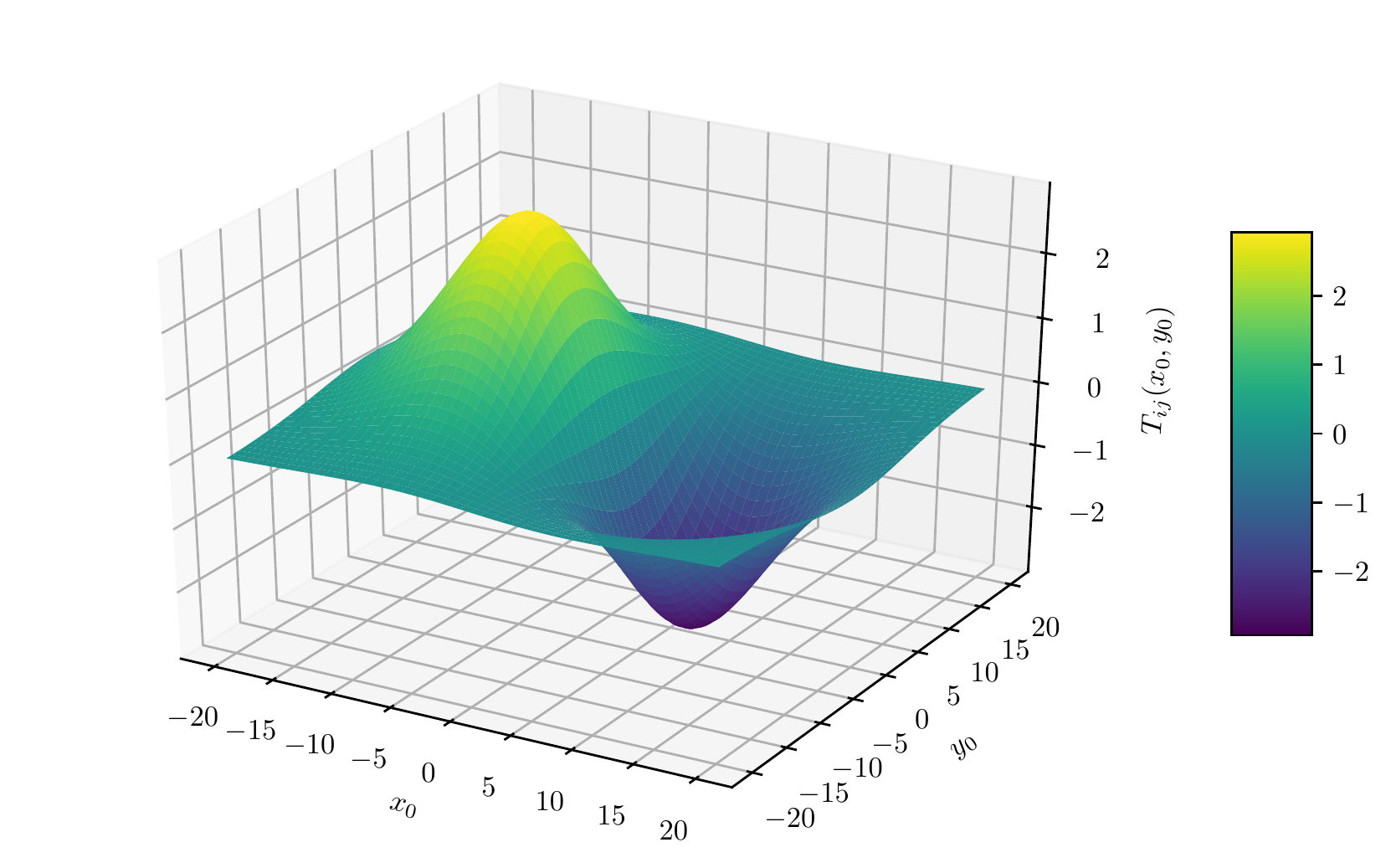}
\end{center}
\caption{Evaluation of $\Im(T_{ij}(r_0))$ for $r_0$ in the $xy$ plane for $G=g=v=1$.}
\label{fig_tijr0}
\end{figure}

As an instructive exercise, taking $G,g,v=1$, we can evaluate this integral. The result is seen in Fig. \ref{fig_tijr0}.  The first thing to note is that the result of integral is symmetric around the $x$-axis, and therefore only dependent on the combination $(y_0^2+z_0^2)$. So, for example, taking $r_0 = (x_0, 2, 3)$ yields the same result as $r_0 = (x_0, \sqrt{13}, 0)$, $r_0 = (x_0, 3, 2)$, etc. In addition, the function is odd under reflection from $x\rightarrow-x$. Therefore, we can evaluate the integral with $z_0 = 0$, $y_0\geq0$, and $x_0\geq0$, without loss of generality. 

\section{Dirac Eigenvalues For Monopole-Antimonopole Ensembles}
\subsection{The monopole ensembles}

We will use the monopole configurations found in our previous path-integral study of monopole Bose-Einstein condensation \cite{Ramamurti:2017fdn}, and study the effects of density and temperature on the eigenvalue spectrum of the Dirac operator for BPS monopoles.

To constrain the values of $g$ and $v$, we can use the mass of the monopole. In the BPS limit, the mass is given by
\eq{
M = \frac{4 \pi v}{g}\,.
}

The mass of the monopole was studied in Ref. \cite{D'Alessandro:2010xg} through lattice simulations of SU(2) pure-gauge theory, which shows a mass of around $2T_c$ at $T=T_c$, and then rapidly grows as temperature increases. The density of monopoles in SU(2) gauge theories was studied in Ref. \cite{DAlessandro:2007lae}, and was parameterized by
\eq{
\frac{\rho}{T_c^3}\left(\frac{T}{T_c}\right) = \frac{0.557 (T/T_c)^3}{\log(2.69 (T/T_c) )^2}\,.
}

\onecolumngrid

\begin{figure}[h!]
\begin{minipage}{\textwidth}
\centering
\subcaptionbox{ $T = 1T_c$}{\includegraphics[width=.48\linewidth]{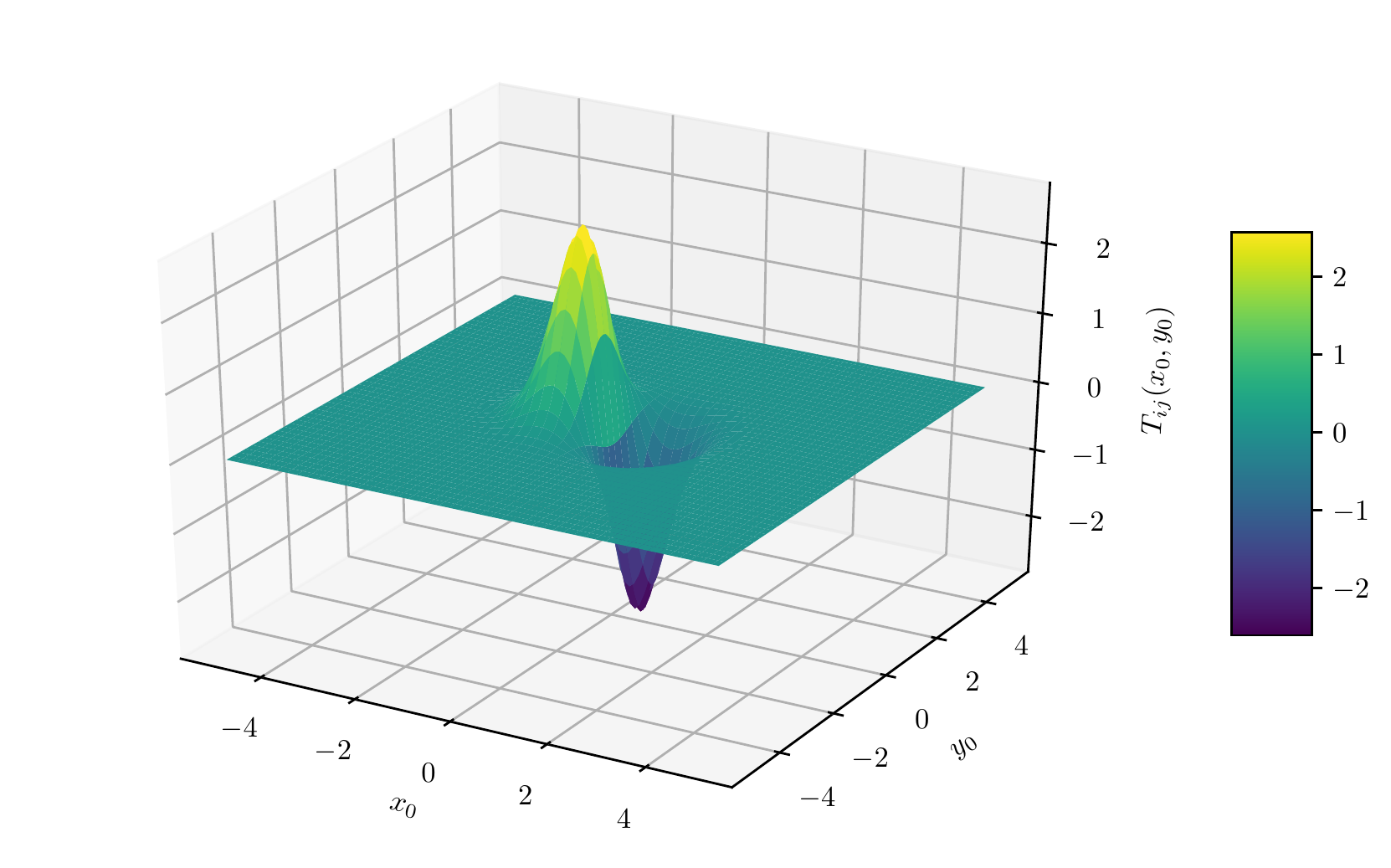}}
\subcaptionbox{$T = 1.05T_c$}{\includegraphics[width=.48\linewidth]{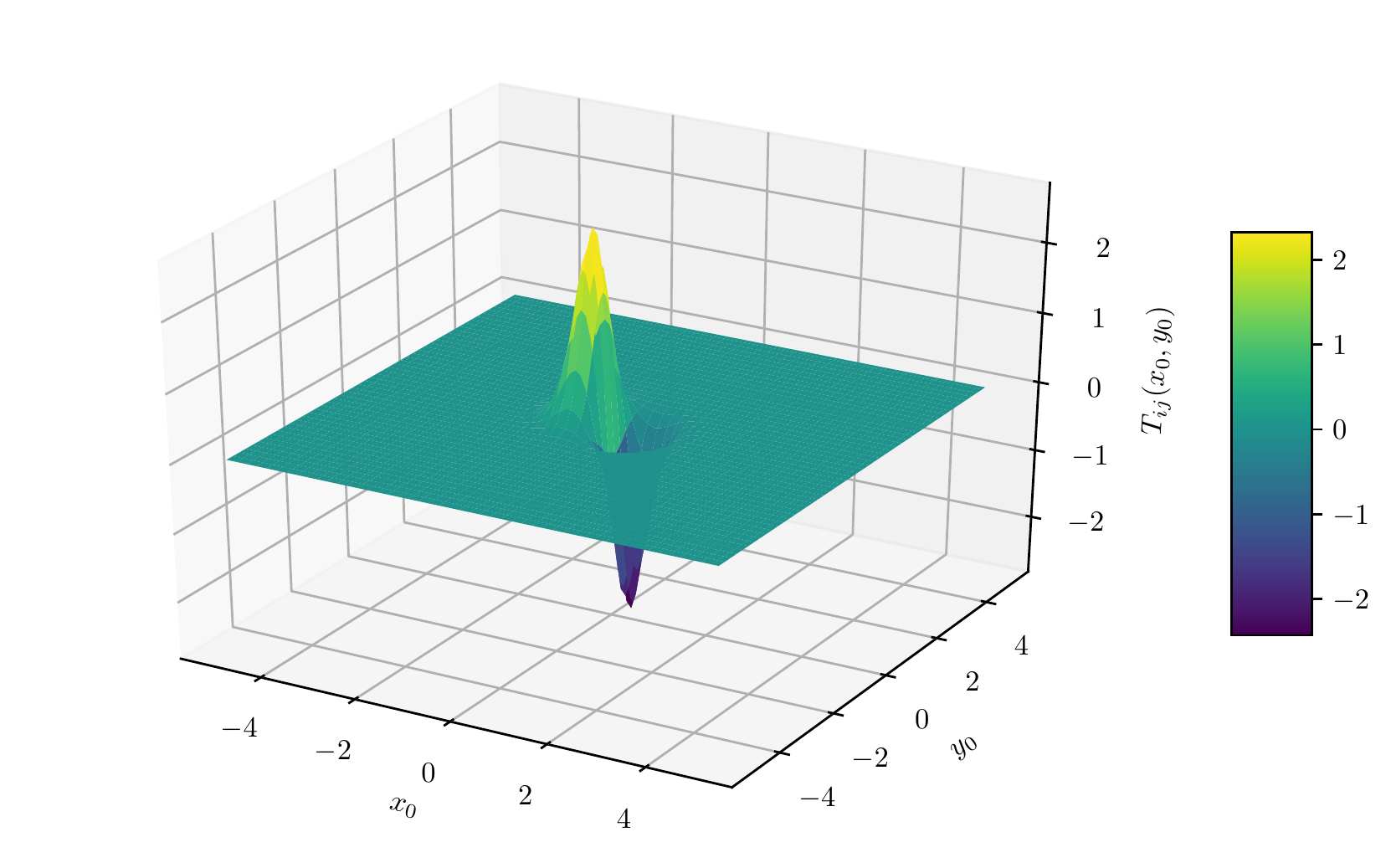}}

\subcaptionbox{$T = 1.1T_c$}{\includegraphics[width=0.48\linewidth]{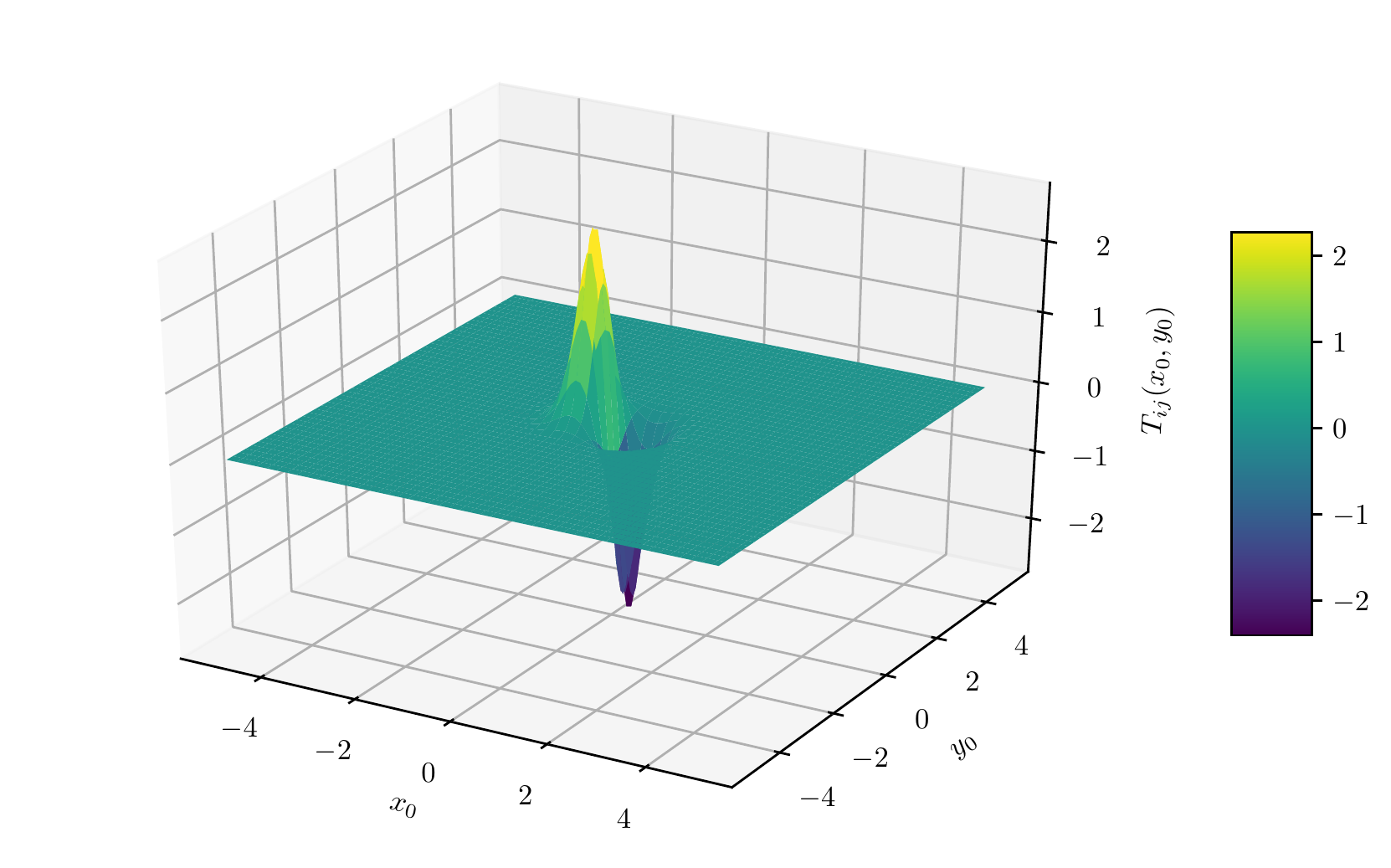}}
\subcaptionbox{$T = 1.2T_c$}{\includegraphics[width=0.48\linewidth]{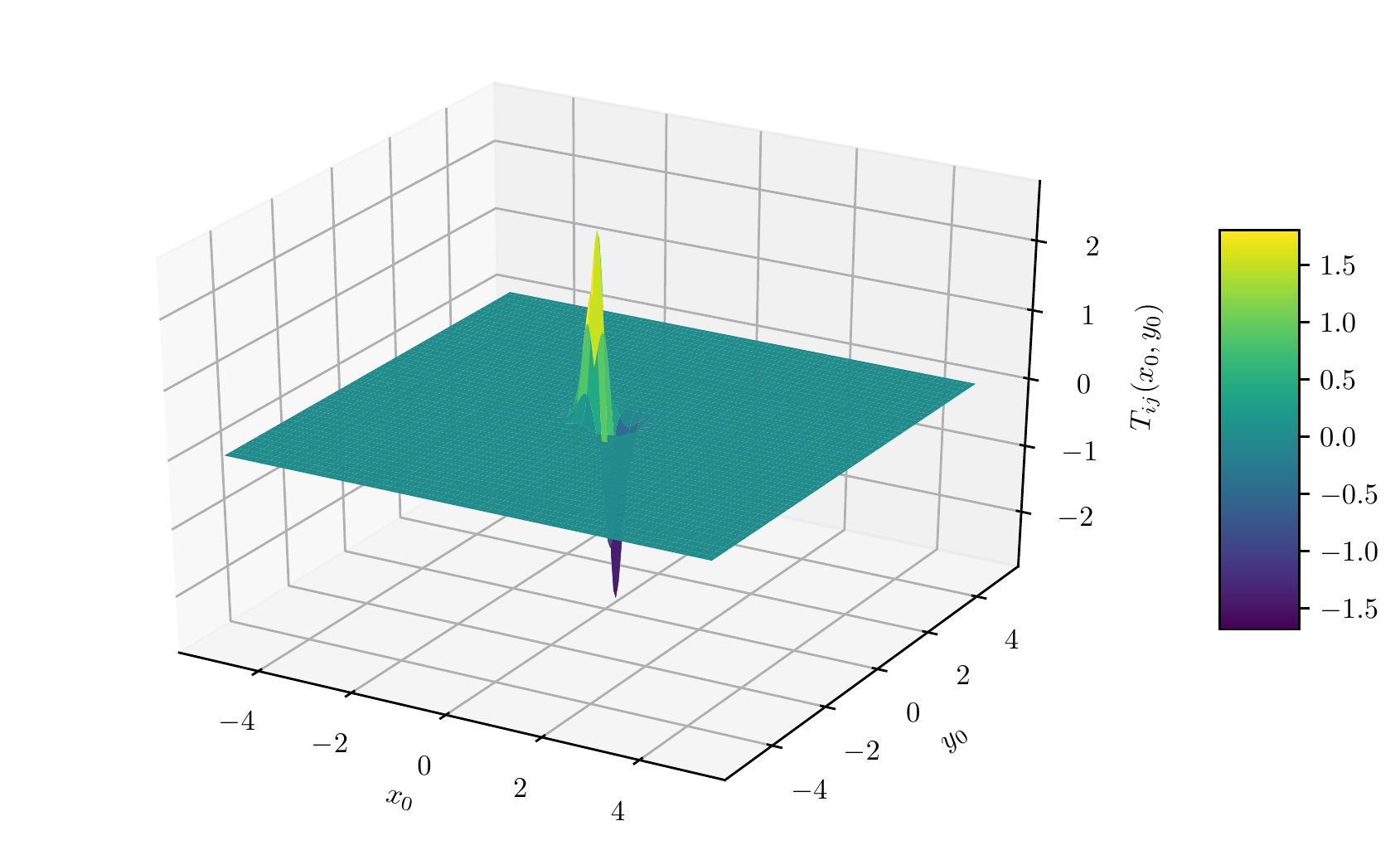}}
\caption{Evaluation of $\Im(T_{ij}(r_0))$ for $r_0$ in the $xy$ plane for different temperatures.}
\label{fig_tij_t}
\end{minipage}
\end{figure}

\twocolumngrid

\subsection{The quantization procedure}

We will evaluate the {\em evolution matrix} $U$, defined as time-ordered integral of the hopping matrix in the previous section over the Matsubara periodic time $\tau\in [0,\beta]$. This matrix will then be diagonalized to find the eigenvalues for the fermion states. Because each eigenstate is still fermionic, each is required to fulfill the fermionic boundary conditions, namely that the state must return to minus itself after one rotation around the Matsubara circle. This defines quantization of the Dirac eigenvalues  by,
\eq{
\lambda_i+\omega_{i,n} = \left(n+\frac{1}{2}\right) \frac{2 \pi}{\beta}\,,
}
where $\lambda_i$s are the eigenvalues of the hopping matrix $\bf{T}$.

For monopoles that move in Euclidean time, we must integrate over the Matsubara circle to find the fermion frequencies,
\eq{
U=\oint_\beta d \tau e^{i H \tau} = -\mathbbm{1} \,.
}
This can be approximated by
\al{
-\mathbbm{1} &\approx  e^{i H_m \Delta \tau}\ldots e^{i H_2 \Delta \tau} e^{i H_1 \Delta \tau} \nn\\
&\approx  (1+ i H_m \Delta \tau-\ldots)\ldots(1+ i H_1 \Delta \tau-\ldots) \nn
}
for $m$ time slices. We diagonalize the resulting matrix on the right-hand side and solve to find the quantity $\lambda + \omega$.

\subsection{Dirac eigenvalue spectra and chiral symmetry breaking}

\begin{figure}[h!]
\begin{center}
\includegraphics[width=.5\textwidth]{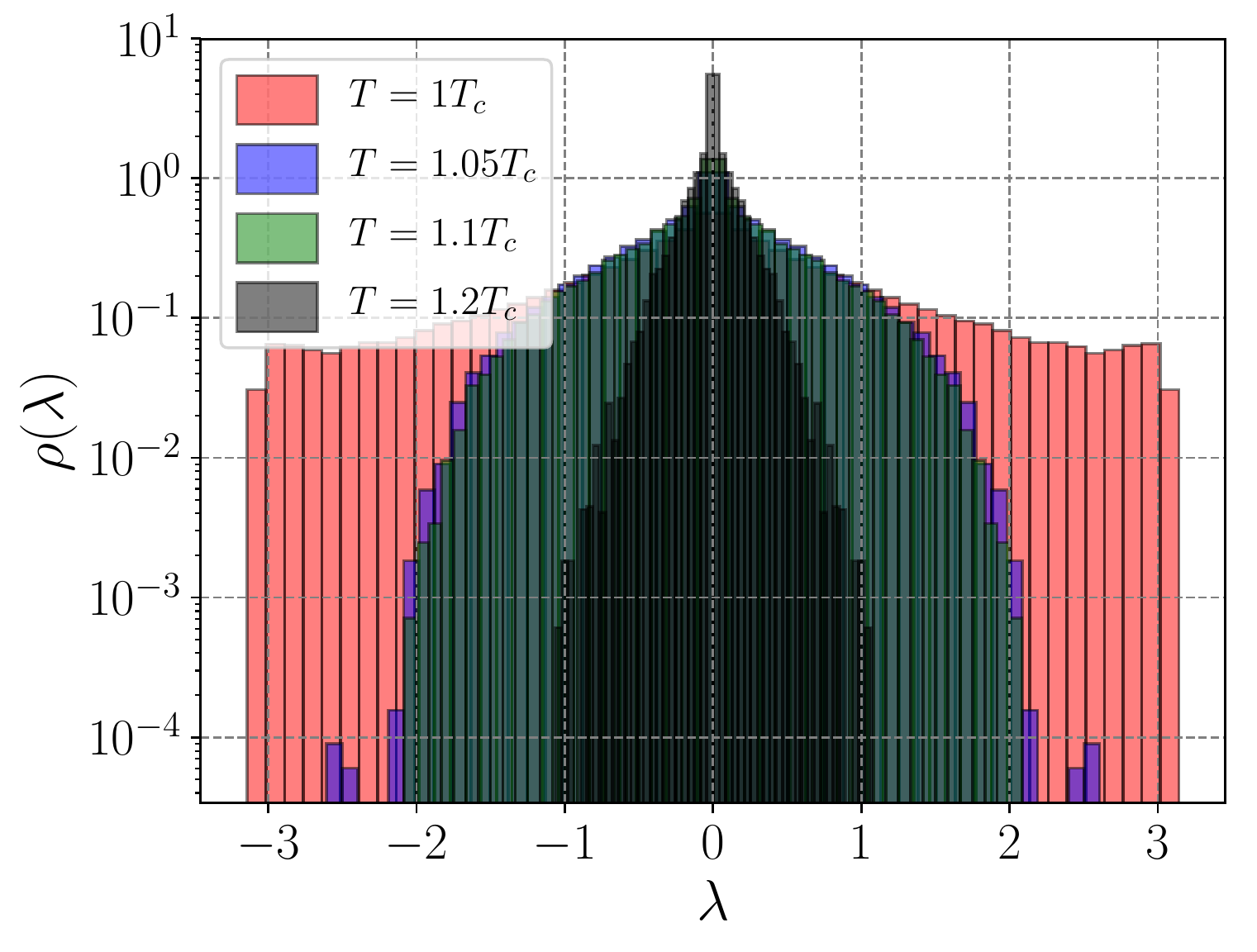}
\end{center}
\caption{Eigenvalue distribution for $T/T_c = 1$ (red),  $1.1\text{ (blue), and } 1.2\text{ (green)}$. Note the logarithmic scale.}
\label{fig_lambdas}
\end{figure}

For simplicity, we will work in units of $T_c$ (i.e. $T_c = 1$) when doing this calculation (for mass and temperature, for example), and units of length will be defined by the density of monopoles $r \sim \rho^ {-1/3}$ in units of $1/T_c$. 

Before we begin, to estimate what the effects of temperature will be on our results, we can evaluate the integral in Eq. (\ref{eq_tij}) with different values of the parameters, corresponding to the range of temperatures we will simulate. We will take $G=g=\sqrt{4\pi/3}$ -- as would be the case in QCD -- and $v$ from the monopole mass, taken from Ref. \cite{D'Alessandro:2010xg}. The results for $T/T_c=1,\, 1.05,\,1.1, \text{ and } 1.2$  are seen in Fig. \ref{fig_tij_t} (a), (b), (c), and (d), respectively.

Compared to the $g=G=v=1$ case, the range of the zero-mode hopping is significantly smaller -- $\sim$10 units of length in $x_0$ to a peak for the former case and 2.5 for $T=T_c$ -- and decreases rapidly with temperature. In addition, the peaks of the function reduce in magnitude and very quickly become much sharper as temperature is increased. Therefore, the contribution to the $T_{ij}$ matrix at temperatures above $T_c$ will only come from ``local'' hopping (i.e. only when there is a monopole-antimonopole molecular bound state), while at $T_c$, the {\em ensemble} contributes and there can be a chain of hopping.

To make this explicit, we construct the hopping matrix, as described above, for SU(2) monopole configurations at $T/T_c=1,\, 1.05, \,1.1,\,1.2 \text{ and } 1.5$, and calculate the eigenvalue spectra of the Dirac operator at each of those temperatures. For each of the temperatures, we take 400 path configurations of 32 time slices, each with 32 particles in a box with periodic boundary conditions -- the box is repeated to the extent necessary to study the whole range of $x_0,y_0,z_0$. At each time slice, the hopping matrix is calculated; the matrices are then path-exponentiated to get the hopping matrix for the ensemble configuration.

\begin{figure*}[th!]
\centering
\subcaptionbox{ $T = 1T_c$ \label{fig_omega1tc}}{\includegraphics[width=.48\linewidth]{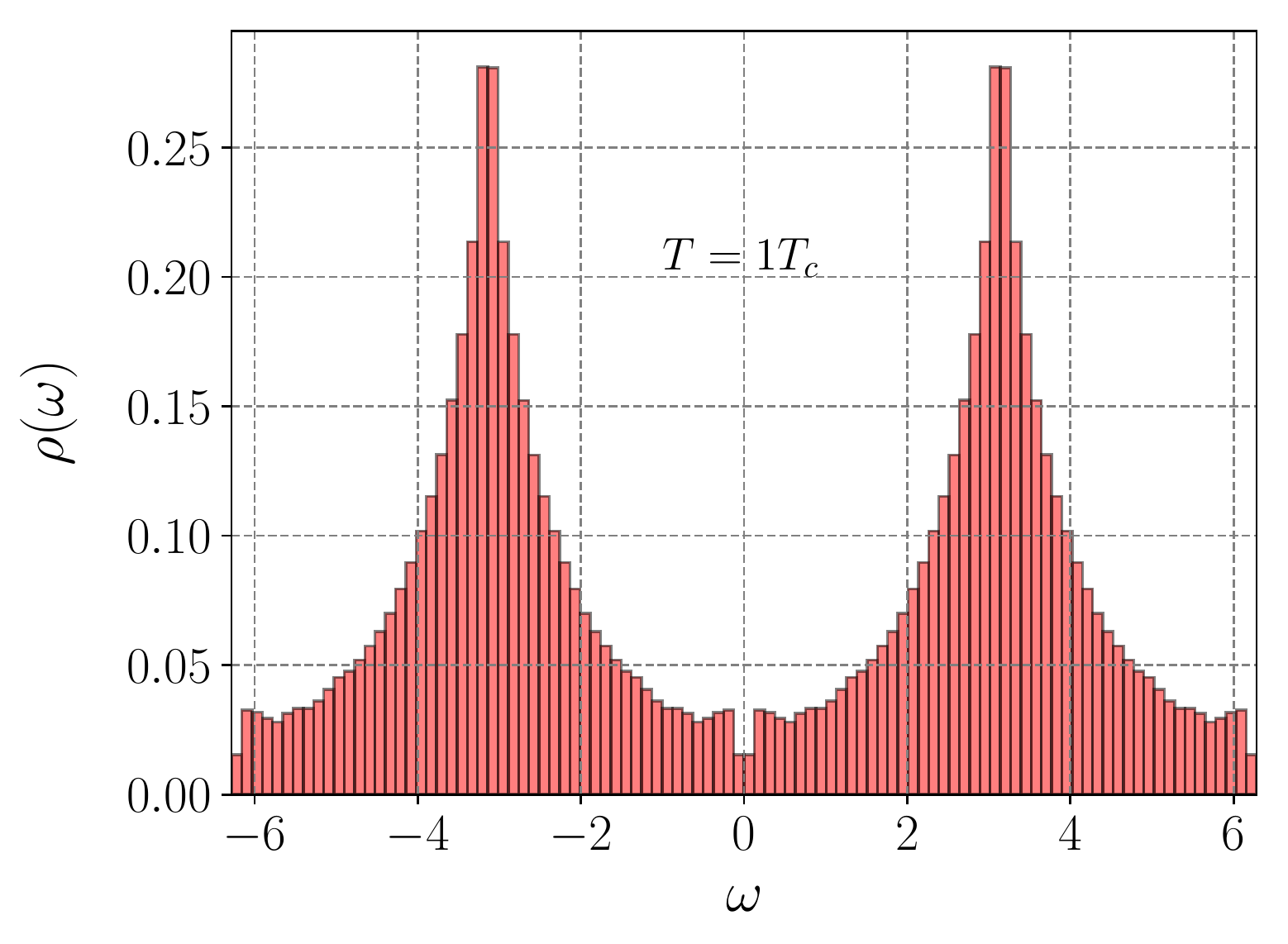} }
\subcaptionbox{$T = 1.05T_c$}{\includegraphics[width=.48\linewidth]{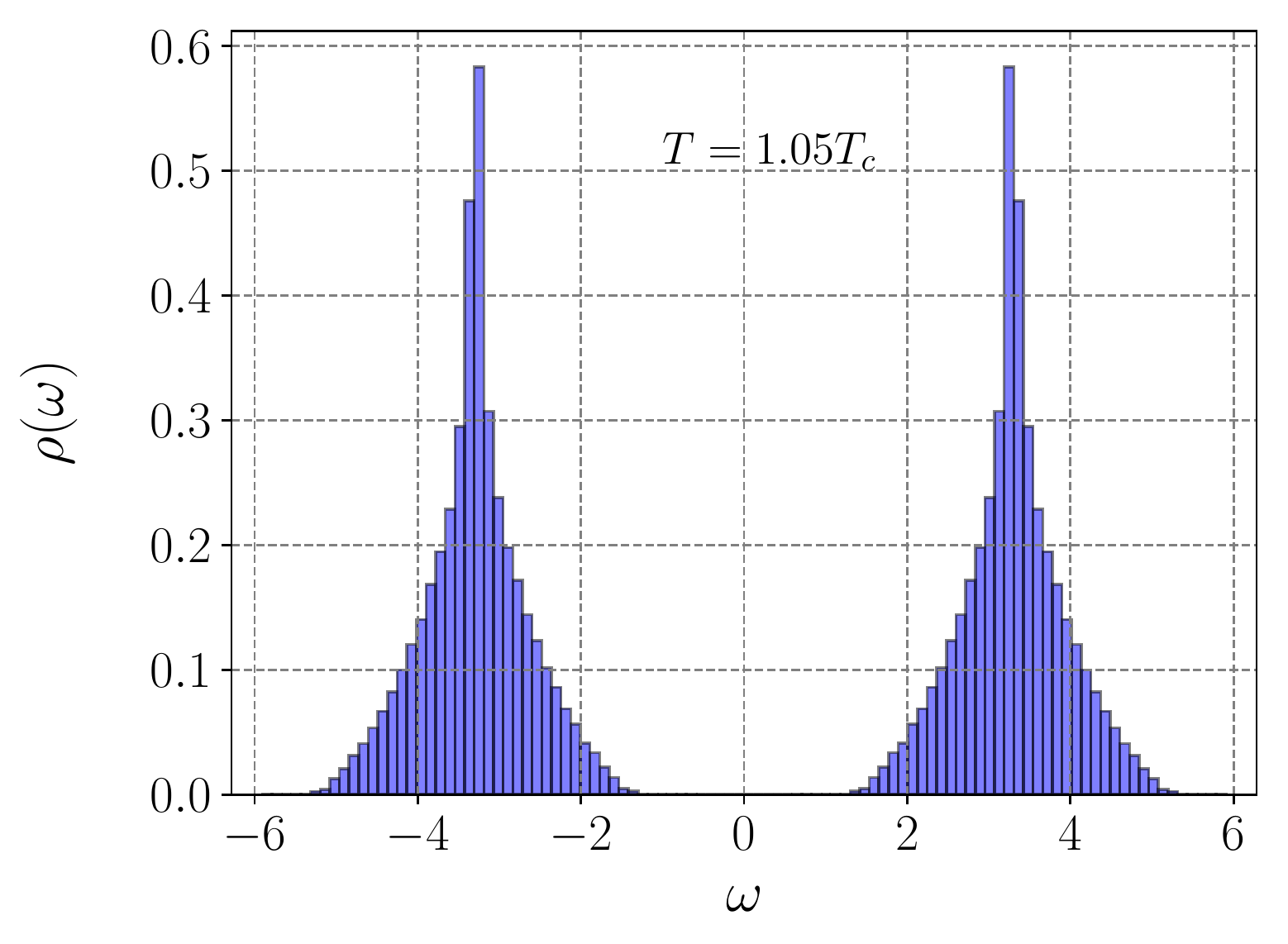}}

\subcaptionbox{$T = 1.1T_c$}{\includegraphics[width=0.48\linewidth]{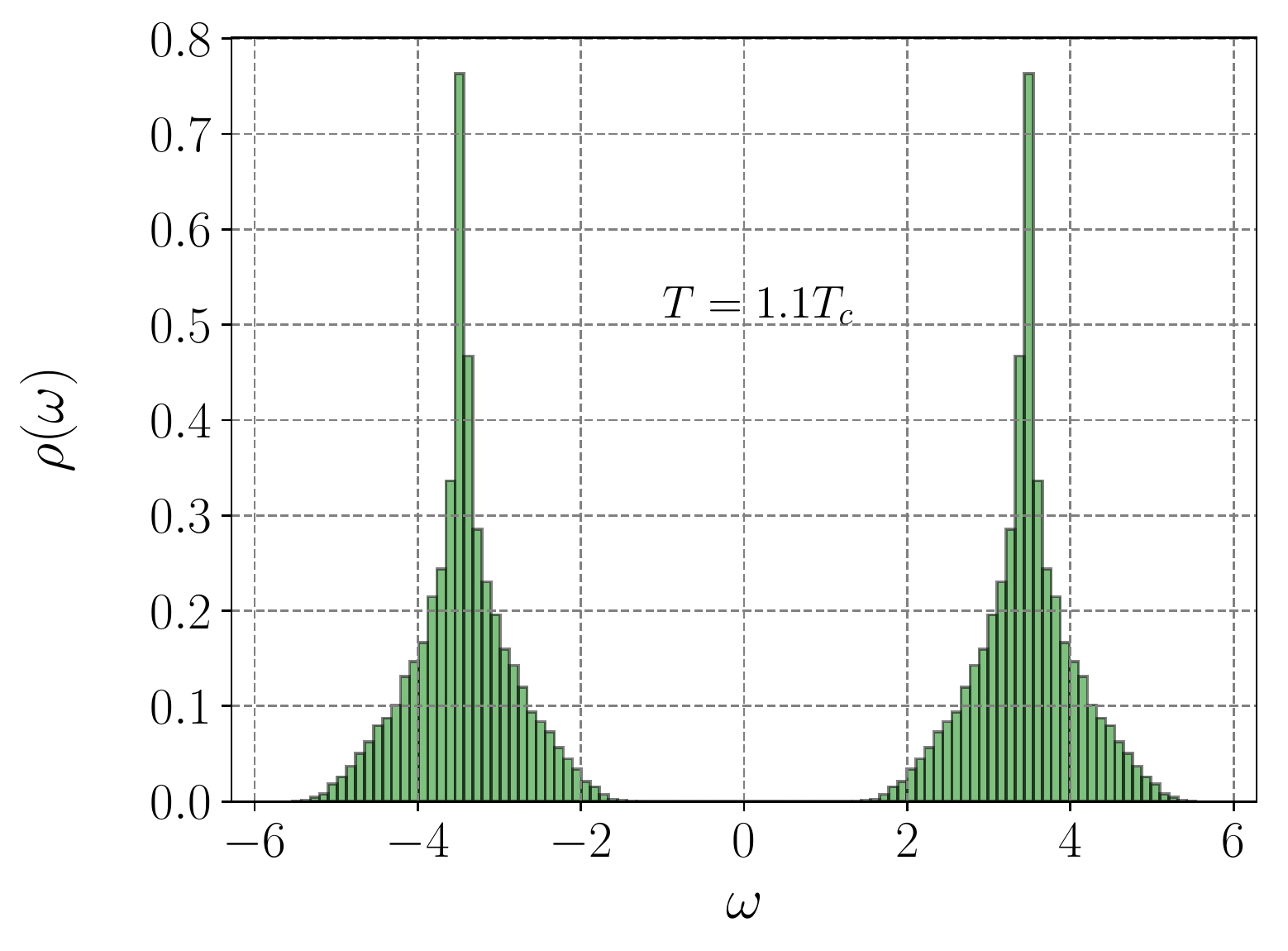}}
\subcaptionbox{$T = 1.2T_c$}{\includegraphics[width=0.48\linewidth]{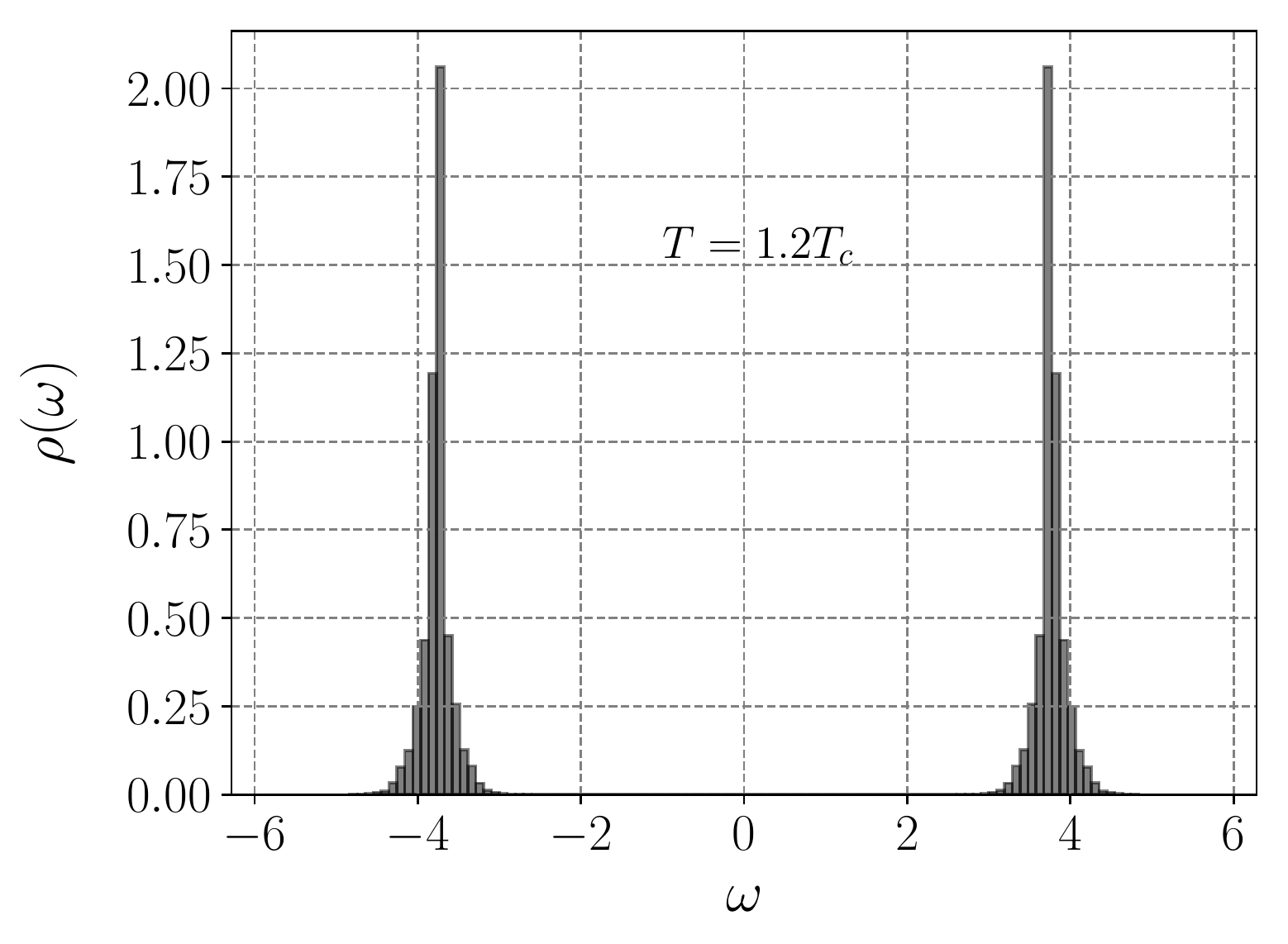}}
\caption{Distributions of Dirac eigenvalues for $T/T_c = $ (a) 1 , (b) 1.05 , (c) 1.1, and (d) 1.2, respectively.}
\label{fig_omegas}
\end{figure*}

The spectral density of eigenvalues is given by
\eq{
\rho(\lambda) = \frac{1}{V} \sum_i \delta(\lambda - \lambda_i)\,.
}
The eigenvalues of the configurations -- before imposing the fermion boundary conditions -- are seen in Fig. \ref{fig_lambdas}. We can then compute the eigenvalues of the Dirac operator with 
\eq{\omega_{i,n} = \left(n+\frac{1}{2}\right) \frac{2 \pi}{\beta} -  \lambda_i \,.}
Considering only the $n=\pm1$ case, so that $\omega_i = \pm \pi T - \lambda_i$, we get the distributions shown in Fig. \ref{fig_omegas} (a), (b), (c), and (d) for $T/T_c = 1,\, 1.05,\, 1.1, \text{ and } 1.2$, respectively. 

\begin{figure}[h!]
\begin{center}
\includegraphics[width=.5\textwidth]{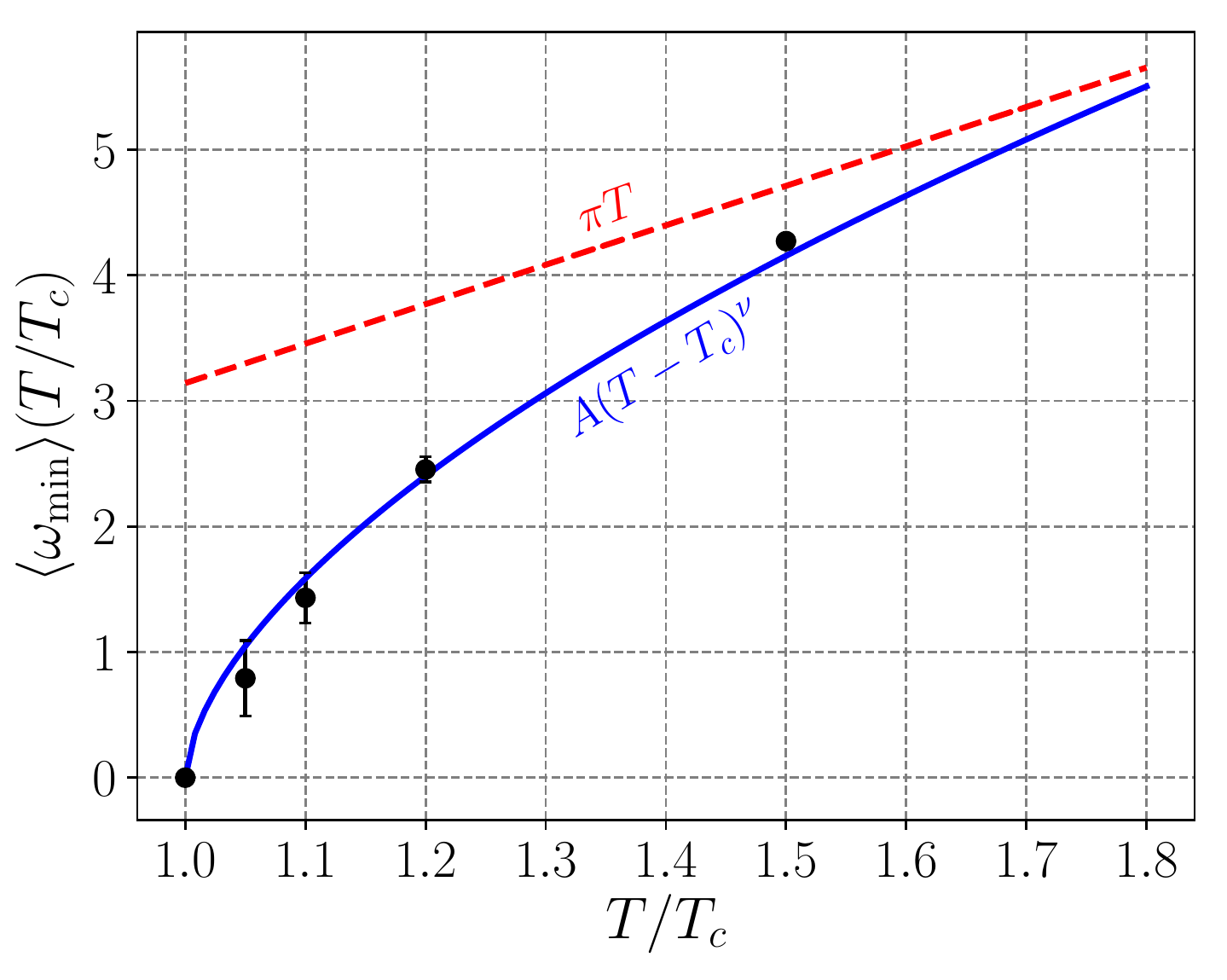}
\end{center}
\caption{Minimum Dirac eigenvalue as a function of temperature. The black dots are values from our simulations, the blue line is the fit $A(T-T_c)^\nu$, and the red dashed line is $\pi T$.}
\label{fig_minomegas}
\end{figure}

The Banks-Casher relation \cite{Banks:1979yr} makes the connection between the density of eigenvalues at $\omega = 0$ and the magnitude of the chiral condensate. In studying the monopole contribution to the chiral condensate, it is important to note that we can only approach the critical temperature $T_c$ from above, as we do not have any lattice data on monopole density, mass, or correlations below $T_c$. As a result, we will primarily focus on the gap in eigenvalue spectrum around $\omega = 0$ as a proxy for the chiral transition.

The first, and most important, thing to notice is that when $T=T_c$, the eigenvalue distribution has a  finite density at $\omega=0$ (see Fig. \ref{fig_omega1tc}), which indicates the nonzero value of the chiral condensate; there is no gap in the spectrum present at $T=T_c$. (A small dip seen around zero is a consequence of finite size of the system,  well known and studied on the lattice and in topological models. It should be essentially ignored in extrapolation to zero.)

Furthermore, one can see the onset of non-zero density at zero eigenvalue by looking at the smallest eigenvalue in each configuration. Fig. \ref{fig_minomegas} shows the mean smallest eigenvalue as a function of temperature. We fit the data with the function
\eq{
\langle \omega_\text{min} \rangle = A (T-T_c)^\nu\,,
}
where $\langle \ldots \rangle$ indicates an average over configurations. The fit parameters were found to be $A = 6.29$ and $\nu = .60$.  We note that this exponent $\nu$ is compatible with the critical exponent found from the diverging correlation length at the deconfinement temperature, which for the 3d Ising model is $\nu \approx .63$, and is also consistent with the monopole BEC transition found in \cite{D'Alessandro:2010xg}. This agreement was in fact not expected, because of quenched nature of the calculation, but has been observed nevertheless. 

\section{Summary}

In qualitative terms, the mechanism of chiral symmetry breaking based on monopoles is as follows. A single monopole (or anti-monopole) generates additional quark and antiquark bound states. At high temperatures, the monopoles have large mass and the probability of hopping is therefore low. The 4d Dirac operator eigenvalues are well localized near the fermionic Matsubara frequencies $2\pi T(n+1/2)$. Using the condensed matter analogy, one may say that a matter is an insulator.

However, as $T$ decreases toward $T_c$, the amplitudes of quark ``hopping" from one monopole to an antimonopole (and vice versa) grow. Eventually, at some critical density, quarks become ``collectivized" and are able to travel very far from their original locations. The physics of the mechanism is similar to insulator-metal transition in condensed matter under pressure.

Technically, the central point is the distinction between the evolution operator and quantization of the fermionic states
on one hand, and the Dirac operator and its eigenvalues. 

Quantitatively, we found that not only the mechanism works in principle, but that a noticeable quark condensate does appear at $T\approx T_c$, practically simultaneously with the deconfinement phase transition, seen by the BEC of monopoles. This observation is consistent with what what has been observed in quenched lattice calculations.

Finally, let us comment on the dependence of chiral symmetry breaking on the fermion periodicity phase. We have not studied it in this work, but note that for {\em periodic} quarks, the Matsubara frequencies shift to bosonic set $2\pi T n$, including $n=0$. Therefore the monopoles would produce a non-zero quark condensate at {\em any} density. This comment implies that the chiral transition is in general some function of the periodicity phase, and its coincidence with deconfinement only happens for the anti-periodic quarks we studied. 
 
\vspace{2ex}
 
{\bf Acknowledgements.}
The authors thank the Institute for Advanced Computational Science (IACS) at Stony Brook University for the use of its LI-red computational cluster. This work was supported in part by the U.S. D.O.E. Office of Science,  under Contract No. DE-FG-88ER40388.

 \appendix 
 
 \section{Gamma matrices and chirality of the monopole zero modes} \label{app_a}
 
 The representation of the Dirac matrices  used by Jackiw and Rebbi 
 and mentioned in the text  correspond to the definition 
 \eq{
 \gamma_4 = \beta \,,\hspace{1cm} \gamma_4 \vec{\gamma} = -i \vec{\alpha}\,,
 }
with the representation of the gamma matrices
    \begin{align}
    \gamma_4  =  -i \begin{pmatrix}
          0 & \mathbbm{1}\\
           -\mathbbm{1} & 0\\
         \end{pmatrix} \,,    \hspace{.3cm} 
    \gamma_i  =   \begin{pmatrix}
          -\sigma_i  & 0\\
          0 & \sigma_i\\
         \end{pmatrix} \,,
         \hspace{.3cm} 
          \gamma_5  =  i \begin{pmatrix}
          0 & \mathbbm{1}\\
           \mathbbm{1} & 0\\
         \end{pmatrix} \,.
  \end{align}

Note that this form is different both from  the standard Dirac representation
    \begin{align}
    \gamma_0  =  \begin{pmatrix}
           \mathbbm{1} & 0\\
           0 &  -\mathbbm{1} \\
         \end{pmatrix}  \,,
         \hspace{.3cm} 
          \gamma_i  =  \begin{pmatrix}
          0 & \sigma_i\\
           -\sigma_i & 0\\
         \end{pmatrix} \,,    \hspace{.3cm} 
         \gamma_5  =  \begin{pmatrix}
          0 & \mathbbm{1}\\
           \mathbbm{1} & 0\\
         \end{pmatrix}\,,
  \end{align}
  and  the Weyl one, in which 
    \begin{align}
    \gamma_0  =  \begin{pmatrix}
            0 & \mathbbm{1}\\
          \mathbbm{1}  & 0\\
         \end{pmatrix}  \,,
         \hspace{.3cm} 
          \gamma_i  =  \begin{pmatrix}
          0 & \sigma_i\\
           -\sigma_i & 0\\
         \end{pmatrix} \,,    \hspace{.3cm} 
         \gamma_5  =  \begin{pmatrix}
         -\mathbbm{1} &0 \\
           0& \mathbbm{1}\\
         \end{pmatrix}\,.
  \end{align}
  
  Standard definition of fermion chirality (left and right polarizations) is related to projectors $(1\pm \gamma_5)/2$, and so only the last Weyl representation, in which $\gamma_5$ 
  is diagonal, is really chiral. The zero modes of pure gauge solitons, such as instantons and instanton-dyons, are chiral in this standard sense.
   
 The  2-component zero modes found by Jackiw and Rebbi are  often called ``chiral" in literature, but they are {\em not} chiral in the standard  sense,  as seen already from the fact that in their representation  $\gamma_5$ is {\em not} diagonal. Furthermore, as seen directly from the Lagrangian of the Georgi-Glashow model,  fermions interact with a scalar field, and this vertex mixes the left and right polarizations, explicitly breaking chiral symmetry.  
  
In pure gauge theory with massless fermions (which we discuss),  the SU$(N_f)$ chiral symmetry is exact. So, when the lattice monopoles --  whatever their microscopic structure may be -- have fermionic bound states, those should belong to the representation of the standard chiral symmetry, rather than the one with the quotation marks. So, while we use the 't Hooft-Polyakov monopole and its Jackiw-Rebbi zero modes as an example, we do not expect it correctly reproduce their chiral properties. We assume that the zero modes of monopoles in gauge theories without  scalars (such as QCD) are truly chiral in the usual sense.


\begin{thebibliography}{9}




\bibitem{Dirac60}
	P. A. M.~Dirac,
	Quantised Singularities in the Electromagnetic Field,
	\href{https://doi.org/10.1098/rspa.1931.0130}{Proc. R. Soc. Lond. A {\bf133} 60-72 (1931)}.
	
%\cite{NUPHA.B79.276}
\bibitem{NUPHA.B79.276} 
  G.~'t Hooft,
  Magnetic Monopoles in Unified Gauge Theories,
  \doi{Nucl.\ Phys.\ B {\bf 79}, 276 (1974)}{10.1016/0550-3213(74)90486-6}.
  %%CITATION = doi:10.1016/0550-3213(74)90486-6;%%
  %2751 citations counted in INSPIRE as of 18 Jan 2018


%\cite{Polyakov:1974ek}
\bibitem{Polyakov:1974ek} 
  A.~M.~Polyakov,
  Particle Spectrum in the Quantum Field Theory,
  JETP Lett.\  {\bf 20}, 194 (1974)
  [Pisma Zh.\ Eksp.\ Teor.\ Fiz.\  {\bf 20}, 430 (1974)].
  %%CITATION = JTPLA,20,194;%%
  %2170 citations counted in INSPIRE as of 18 Jan 2018


%\cite{Shuryak:2017kct}
\bibitem{Shuryak:2017kct} 
  E.~Shuryak,
  Instanton-dyon ensembles reproduce deconfinement and chiral restoration phase transitions,
  \hri{1710.03611}{hep-lat}.
  %%CITATION = ARXIV:1710.03611;%%


%\cite{Dorey:2000qc}
\bibitem{Dorey:2000qc} 
  N.~Dorey and A.~Parnachev,
  Instantons, compactification and S duality in N=4 SUSY Yang-Mills theory. 2.,
  \doi{JHEP {\bf 0108}, 059 (2001)}{10.1088/1126-6708/2001/08/059},
  \hre{hep-th}{0011202}.
  %%CITATION = doi:10.1088/1126-6708/2001/08/059;%%
  %15 citations counted in INSPIRE as of 18 Jan 2018


%\cite{Poppitz:2011wy}
\bibitem{Poppitz:2011wy} 
  E.~Poppitz and M.~Unsal,
  Seiberg-Witten and 'Polyakov-like' magnetic bion confinements are continuously connected,
  \doi{JHEP {\bf 1107}, 082 (2011)}{10.1007/JHEP07(2011)082},
  \hri{1105.3969}{hep-th}.
  %%CITATION = doi:10.1007/JHEP07(2011)082;%%
  %58 citations counted in INSPIRE as of 18 Jan 2018


%\cite{Poppitz:2012sw}
\bibitem{Poppitz:2012sw} 
  E.~Poppitz, T.~Sch\"{a}fer and M.~Unsal,
  Continuity, Deconfinement, and (Super) Yang-Mills Theory,
  \doi{JHEP {\bf 1210}, 115 (2012)}{10.1007/JHEP10(2012)115},
  \hri{1205.0290}{hep-th}.
  %%CITATION = doi:10.1007/JHEP10(2012)115;%%
  %82 citations counted in INSPIRE as of 18 Jan 2018


%\cite{Nambu:1974zg}
\bibitem{Nambu:1974zg} 
  Y.~Nambu,
  Strings, Monopoles and Gauge Fields,
  \doi{Phys.\ Rev.\ D {\bf 10}, 4262 (1974)}{10.1103/PhysRevD.10.4262}.
  %%CITATION = doi:10.1103/PhysRevD.10.4262;%%
  %931 citations counted in INSPIRE as of 18 Jan 2018


%\cite{NUPHA.B190.455}
\bibitem{NUPHA.B190.455} 
  G.~'t Hooft,
 Topology of the Gauge Condition and New Confinement Phases in Nonabelian Gauge Theories,
  \doi{Nucl.\ Phys.\ B {\bf 190}, 455 (1981)}{10.1016/0550-3213(81)90442-9}.
  %%CITATION = doi:10.1016/0550-3213(81)90442-9;%%
  %1445 citations counted in INSPIRE as of 18 Jan 2018


%\cite{Mandelstam:1974pi}
\bibitem{Mandelstam:1974pi} 
  S.~Mandelstam,
  Vortices and Quark Confinement in Nonabelian Gauge Theories,
  \doi{Phys.\ Rept.\  {\bf 23}, 245 (1976)}{10.1016/0370-1573(76)90043-0}.
  %%CITATION = doi:10.1016/0370-1573(76)90043-0;%%
  %1108 citations counted in INSPIRE as of 18 Jan 2018


%\cite{Laursen:1987eb}
\bibitem{Laursen:1987eb} 
  M.~L.~Laursen and G.~Schierholz,
  Evidence for Monopoles in the Quantized SU(2) Lattice Vacuum: A Study at Finite Temperature,
  \doi{Z.\ Phys.\ C {\bf 38}, 501 (1988)}{10.1007/BF01584402}
  %%CITATION = doi:10.1007/BF01584402;%%
  %25 citations counted in INSPIRE as of 18 Jan 2018


%\cite{Koma:2003gq}
\bibitem{Koma:2003gq} 
  Y.~Koma, M.~Koma, E.~M.~Ilgenfritz, T.~Suzuki and M.~I.~Polikarpov,
  Duality of gauge field singularities and the structure of the flux tube in Abelian projected SU(2) gauge theory and the dual Abelian Higgs model,
  \doi{Phys.\ Rev.\ D {\bf 68}, 094018 (2003)}{10.1103/PhysRevD.68.094018},
  \hre{hep-lat}{0302006}.
  %%CITATION = doi:10.1103/PhysRevD.68.094018;%%
  %42 citations counted in INSPIRE as of 18 Jan 2018


%\cite{Bornyakov:2003vx}
\bibitem{Bornyakov:2003vx} 
  V.~G.~Bornyakov {\it et al.} [DIK Collaboration],
  Dynamics of monopoles and flux tubes in two flavor dynamical QCD,
  \doi{Phys.\ Rev.\ D {\bf 70}, 074511 (2004)}{10.1103/PhysRevD.70.074511},
  \hre{hep-lat}{0310011}.
  %%CITATION = doi:10.1103/PhysRevD.70.074511;%%
  %34 citations counted in INSPIRE as of 18 Jan 2018


%\cite{Suzuki:2009xy}
\bibitem{Suzuki:2009xy} 
  T.~Suzuki, M.~Hasegawa, K.~Ishiguro, Y.~Koma and T.~Sekido,
  Gauge invariance of color confinement due to the dual Meissner effect caused by Abelian monopoles,
  \doi{Phys.\ Rev.\ D {\bf 80}, 054504 (2009)}{10.1103/PhysRevD.80.054504},
  \hri{0907.0583}{hep-lat}.
  %%CITATION = doi:10.1103/PhysRevD.80.054504;%%
  %18 citations counted in INSPIRE as of 18 Jan 2018


%\cite{DAlessandro:2007lae}
\bibitem{DAlessandro:2007lae} 
  A.~D'Alessandro and M.~D'Elia,
  Magnetic monopoles in the high temperature phase of Yang-Mills theories,
  \doi{Nucl.\ Phys.\ B {\bf 799}, 241 (2008)}{10.1016/j.nuclphysb.2008.03.002},
  \hri{0711.1266}{hep-lat}.
  %%CITATION = doi:10.1016/j.nuclphysb.2008.03.002;%%
  %66 citations counted in INSPIRE as of 18 Jan 2018


%\cite{Bonati:2013bga}
\bibitem{Bonati:2013bga} 
  C.~Bonati and M.~D'Elia,
  The Maximal Abelian Gauge in SU(N) gauge theories and thermal monopoles for N = 3,
  \doi{Nucl.\ Phys.\ B {\bf 877}, 233 (2013)}{10.1016/j.nuclphysb.2013.10.004},
  \hri{1308.0302}{hep-lat}.
  %%CITATION = doi:10.1016/j.nuclphysb.2013.10.004;%%
  %16 citations counted in INSPIRE as of 18 Jan 2018


%\cite{Liao:2008jg}
\bibitem{Liao:2008jg} 
  J.~Liao and E.~Shuryak,
  Magnetic Component of Quark-Gluon Plasma is also a Liquid!,
  \doi{Phys.\ Rev.\ Lett.\  {\bf 101}, 162302 (2008)}{10.1103/PhysRevLett.101.162302},
  \hri{0804.0255}{hep-ph}.
  %%CITATION = doi:10.1103/PhysRevLett.101.162302;%%
  %99 citations counted in INSPIRE as of 18 Jan 2018


%\cite{D'Alessandro:2010xg}
\bibitem{D'Alessandro:2010xg} 
  A.~D'Alessandro, M.~D'Elia and E.~V.~Shuryak,
  Thermal Monopole Condensation and Confinement in finite temperature Yang-Mills Theories,
  \doi{Phys.\ Rev.\ D {\bf 81}, 094501 (2010)}{10.1103/PhysRevD.81.094501},
  \hri{1002.4161}{hep-lat}.
  %%CITATION = doi:10.1103/PhysRevD.81.094501;%%
  %43 citations counted in INSPIRE as of 18 Jan 2018


%\cite{Ramamurti:2017fdn}
\bibitem{Ramamurti:2017fdn} 
  A.~Ramamurti and E.~Shuryak,
  Effective Model of QCD Magnetic Monopoles From Numerical Study of One- and Two-Component Coulomb Quantum Bose Gases,
  \doi{Phys.\ Rev.\ D {\bf 95}, no. 7, 076019 (2017)}{10.1103/PhysRevD.95.076019},
  \hri{1702.07723}{hep-ph}.
  %%CITATION = doi:10.1103/PhysRevD.95.076019;%%
  %3 citations counted in INSPIRE as of 18 Jan 2018


%\cite{Bonati:2010tz}
\bibitem{Bonati:2010tz} 
  C.~Bonati, A.~Di Giacomo, L.~Lepori and F.~Pucci,
  Monopoles, abelian projection and gauge invariance,
  \doi{Phys.\ Rev.\ D {\bf 81}, 085022 (2010)}{10.1103/PhysRevD.81.085022},
  \hri{1002.3874}{hep-lat}.
  %%CITATION = doi:10.1103/PhysRevD.81.085022;%%
  %20 citations counted in INSPIRE as of 18 Jan 2018


%\cite{Bonati:2010bb}
\bibitem{Bonati:2010bb} 
  C.~Bonati, A.~Di Giacomo and M.~D'Elia,
  Detecting monopoles on the lattice,
  \doi{Phys.\ Rev.\ D {\bf 82}, 094509 (2010)}{10.1103/PhysRevD.82.094509},
  \hri{1009.2425}{hep-lat}.
  %%CITATION = doi:10.1103/PhysRevD.82.094509;%%
  %11 citations counted in INSPIRE as of 18 Jan 2018


%\cite{DiGiacomo:2017blx}
\bibitem{DiGiacomo:2017blx} 
  A.~Di Giacomo,
  QCD monopoles, abelian projections and gauge invariance,
  \hri{1707.07896}{hep-lat}.
  %%CITATION = ARXIV:1707.07896;%%


%\cite{Liao:2006ry}
\bibitem{Liao:2006ry} 
  J.~Liao and E.~Shuryak,
  Strongly coupled plasma with electric and magnetic charges,
  \doi{Phys.\ Rev.\ C {\bf 75}, 054907 (2007)}{10.1103/PhysRevC.75.054907},
  \hre{hep-ph}{0611131}.
  %%CITATION = doi:10.1103/PhysRevC.75.054907;%%
  %134 citations counted in INSPIRE as of 18 Jan 2018


%\cite{Liao:2007mj}
\bibitem{Liao:2007mj} 
  J.~Liao and E.~Shuryak,
  Electric Flux Tube in Magnetic Plasma,
  \doi{Phys.\ Rev.\ C {\bf 77}, 064905 (2008)}{10.1103/PhysRevC.77.064905},
 \hri{0706.4465}{hep-ph}.
  %%CITATION = doi:10.1103/PhysRevC.77.064905;%%
  %43 citations counted in INSPIRE as of 18 Jan 2018


%\cite{Ratti:2008jz}
\bibitem{Ratti:2008jz} 
  C.~Ratti and E.~Shuryak,
  The Role of monopoles in a Gluon Plasma,
  \doi{Phys.\ Rev.\ D {\bf 80}, 034004 (2009)}{10.1103/PhysRevD.80.034004},
  \hri{0811.4174}{hep-ph}.
  %%CITATION = doi:10.1103/PhysRevD.80.034004;%%
  %46 citations counted in INSPIRE as of 18 Jan 2018


%\cite{Xu:2015bbz}
\bibitem{Xu:2015bbz} 
  J.~Xu, J.~Liao and M.~Gyulassy,
  Bridging Soft-Hard Transport Properties of Quark-Gluon Plasmas with CUJET3.0,
  \doi{JHEP {\bf 1602}, 169 (2016)}{10.1007/JHEP02(2016)169},
  \hri{1508.00552}{hep-ph}.
  %%CITATION = doi:10.1007/JHEP02(2016)169;%%
  %41 citations counted in INSPIRE as of 18 Jan 2018


%\cite{Xu:2014tda}
\bibitem{Xu:2014tda} 
  J.~Xu, J.~Liao and M.~Gyulassy,
  Consistency of Perfect Fluidity and Jet Quenching in semi-Quark-Gluon Monopole Plasmas,
  \doi{Chin.\ Phys.\ Lett.\  {\bf 32}, no. 9, 092501 (2015)}{10.1088/0256-307X/32/9/092501},
  \hri{1411.3673}{hep-ph}.
  %%CITATION = doi:10.1088/0256-307X/32/9/092501;%%
  %57 citations counted in INSPIRE as of 18 Jan 2018


%\cite{Ramamurti:2017zjn}
\bibitem{Ramamurti:2017zjn} 
  A.~Ramamurti and E.~Shuryak,
  The Role of QCD Monopoles in Jet Quenching,
  \doi{Phys.\ Rev.\ D {\bf 97}, 016010 (2018)}{10.1103/PhysRevD.97.016010},
  \hri{1708.04254}{hep-ph}.
  %%CITATION = doi:10.1103/PhysRevD.97.016010;%%
  %2 citations counted in INSPIRE as of 21 Jan 2018

%\cite{Cea:2017ocq}
\bibitem{Cea:2017ocq} 
  P.~Cea, L.~Cosmai, F.~Cuteri and A.~Papa,
  Flux tubes in the QCD vacuum,
  \doi{Phys.\ Rev.\ D {\bf 95}, no. 11, 114511 (2017)}{10.1103/PhysRevD.95.114511},
  \hri{1702.06437}{hep-lat}.
  %%CITATION = doi:10.1103/PhysRevD.95.114511;%%
  %3 citations counted in INSPIRE as of 18 Jan 2018

%\cite{Suzuki:1989gp}
\bibitem{Suzuki:1989gp} 
  T.~Suzuki and I.~Yotsuyanagi,
  A possible evidence for Abelian dominance in quark confinement,
  \doi{Phys.\ Rev.\ D {\bf 42}, 4257 (1990)}{10.1103/PhysRevD.42.4257}.
  %%CITATION = doi:10.1103/PhysRevD.42.4257;%%
  %395 citations counted in INSPIRE as of 21 Jan 2018
  
%\cite{Miyamura:1995xn}
\bibitem{Miyamura:1995xn} 
  O.~Miyamura,
  Chiral symmetry breaking in gauge fields dominated by monopoles on SU(2) lattices,
  \doi{Phys.\ Lett.\ B {\bf 353}, 91 (1995)}{10.1016/0370-2693(95)00530-X}.
  %%CITATION = doi:10.1016/0370-2693(95)00530-X;%%
  %90 citations counted in INSPIRE as of 21 Jan 2018

%\cite{Larsen:2015tso}
\bibitem{Larsen:2015tso} 
  R.~Larsen and E.~Shuryak,
  Instanton-dyon Ensemble with two Dynamical Quarks: the Chiral Symmetry Breaking,
  \doi{Phys.\ Rev.\ D {\bf 93}, no. 5, 054029 (2016)}{10.1103/PhysRevD.93.054029},
  \hri{1511.02237}{hep-ph}.
  %%CITATION = doi:10.1103/PhysRevD.93.054029;%%
  %22 citations counted in INSPIRE as of 18 Jan 2018


%\cite{Schafer:1996wv}
\bibitem{Schafer:1996wv} 
  T.~Sch\"{a}fer and E.~V.~Shuryak,
  Instantons in QCD,
  \doi{Rev.\ Mod.\ Phys.\  {\bf 70}, 323 (1998)}{10.1103/RevModPhys.70.323},
  \hre{hep-ph}{9610451}.
  %%CITATION = doi:10.1103/RevModPhys.70.323;%%
  %1245 citations counted in INSPIRE as of 18 Jan 2018


%\cite{Banks:1979yr}
\bibitem{Banks:1979yr} 
  T.~Banks and A.~Casher,
  Chiral Symmetry Breaking in Confining Theories,
  \doi{Nucl.\ Phys.\ B {\bf 169}, 103 (1980)}{10.1016/0550-3213(80)90255-2}.
  %%CITATION = doi:10.1016/0550-3213(80)90255-2;%%
  %850 citations counted in INSPIRE as of 18 Jan 2018


%\cite{Jackiw:1975fn}
\bibitem{Jackiw:1975fn} 
  R.~Jackiw and C.~Rebbi,
  Solitons with Fermion Number 1/2,
  \doi{Phys.\ Rev.\ D {\bf 13}, 3398 (1976)}{10.1103/PhysRevD.13.3398}.
  %%CITATION = doi:10.1103/PhysRevD.13.3398;%%
  %1070 citations counted in INSPIRE as of 18 Jan 2018


%\cite{Callias:1977cc}
\bibitem{Callias:1977cc} 
  C.~J.~Callias,
  Spectra of Fermions in Monopole Fields: Exactly Soluble Models,
  \doi{Phys.\ Rev.\ D {\bf 16}, 3068 (1977)}{10.1103/PhysRevD.16.3068}.
  %%CITATION = doi:10.1103/PhysRevD.16.3068;%%
  %56 citations counted in INSPIRE as of 18 Jan 2018


%\cite{Sinha:1976bw}
\bibitem{Sinha:1976bw} 
  A.~Sinha,
  SU(3) Magnetic Monopoles,
  \doi{Phys.\ Rev.\ D {\bf 14}, 2016 (1976)}{10.1103/PhysRevD.14.2016}.
  %%CITATION = doi:10.1103/PhysRevD.14.2016;%%
  %36 citations counted in INSPIRE as of 18 Jan 2018


%\cite{Bogomolny:1975de}
\bibitem{Bogomolny:1975de} 
  E.~B.~Bogomolny,
  Stability of Classical Solutions,
  Sov.\ J.\ Nucl.\ Phys.\  {\bf 24}, 449 (1976)
  [Yad.\ Fiz.\  {\bf 24}, 861 (1976)].
  %%CITATION = SJNCA,24,449;%%
  %1579 citations counted in INSPIRE as of 18 Jan 2018


%\cite{Prasad:1975kr}
\bibitem{Prasad:1975kr} 
  M.~K.~Prasad and C.~M.~Sommerfield,
  An Exact Classical Solution for the 't Hooft Monopole and the Julia-Zee Dyon,
  \doi{Phys.\ Rev.\ Lett.\  {\bf 35}, 760 (1975)}{10.1103/PhysRevLett.35.760}.
  %%CITATION = doi:10.1103/PhysRevLett.35.760;%%
  %1357 citations counted in INSPIRE as of 18 Jan 2018
  
\end{thebibliography}
\end{document}